%% file: template-epjc.tex
\RequirePackage{fix-cm}
\documentclass[twocolumn,epjc3]{svjour3}  
\smartqed  
\RequirePackage{graphicx}
\usepackage{makecell}
\usepackage{url}
\usepackage{siunitx}
\usepackage{upgreek}
\usepackage{hhline}
\usepackage{booktabs}
\usepackage{amssymb}
\usepackage{amsmath}
\usepackage{stfloats}
\usepackage{doi}
\usepackage{float}
\usepackage{hyperref}
\usepackage{orcidlink}
\usepackage{placeins}
\usepackage{threeparttable}

\journalname{Eur. Phys. J. C}
\begin{document}

\title{Measurement of the Muon Flux at SND@LHC
}
\subtitle{Results from the 2023-2025 Proton and Heavy-Ion Periods}


\author{The SND@LHC Collaboration\thanksref{e1}$^\text{\normalfont,1}$}
\thankstext{e1}{e-mail: ivaylo.dionisov@cern.ch}
\institute{Full authorlist at the end of the article.\label{1}}



\date{Received: date / Accepted: date}
\maketitle

\begin{abstract}
The SND@LHC experiment investigates neutrinos in the $7.2 < \eta < 8.4$ forward pseudorapidity range. The detector consists of a veto system, a scintillating fiber tracker interleaved with emulsion cloud chambers, and a downstream muon system. Muons originating from collisions at ATLAS (IP1) constitute the primary background for CC neutrino interactions and determine the replacement frequency of the emulsion target. A precise characterization of this flux is therefore essential.

In this work, we report the muon flux measured in the central $31 \times 31$ $\mathrm{cm^2}$ fiducial area of the detector using data from 2023 through 2025. The measured fluxes for \textbf{proton collisions} are:
\num{1.90(4)e-2}~\unit{nb/cm^2} (2023), 
\num{3.76(9)e-2}~\unit{nb/cm^2} (2024), and \\
\num{2.48(5)e-2}~\unit{nb/cm^2} (2025). A 2024 reference proton run at $\sqrt{s} = 5.36\text{ TeV}$ yielded \\\mbox{\num{4.21(14)e-2}~\unit{nb/cm^2}}, providing a direct baseline for the heavy-ion energy regime.

The measured fluxes for \textbf{heavy-ion collisions} are 
\num{3.11(12)e4}~\unit{nb/cm^2},
\num{5.53(22)e4}~\unit{nb/cm^2}, and
\num{3.24(13)e4}~\unit{nb/cm^2} 
in 2023, 2024, and 2025, respectively. 

Uncertainties are dominated by systematic effects, with the statistical component contributing $\lesssim$0.1\% to the total uncertainty.

These results are in agreement with Monte Carlo predictions.

\keywords{Muon flux \and SND@LHC \and CERN \and LHC \and Muon background}
\end{abstract}


\section{Introduction}
\label{intro}
The SND@LHC experiment \cite{bib:sndlhc} is designed to detect high-energy neutrinos in the forward pseudorapidity region $7.2 < \eta < 8.4$ produced by proton or ion collisions at IP1. The vast majority of observed events are from passing muons, which constitute the primary background to neutrino searches. Moreover, preventing excessive exposure of the emulsion films requires regular target replacement. The frequency of these replacements is directly dependent on the muon rate. Additionally, heavy-ion data further provides suitable conditions for studying detector response and tracking performance for low-energy muons. For these reasons, a precise measurement of the muon flux is essential.

We previously reported corresponding measurements for proton collisions in 2022 \cite{bib:ppMuonFlux}. This paper gives an update for proton collisions and additional measurements for heavy-ion collisions in 2023, 2024 and 2025.

The muon flux is given by:
\begin{equation}
    \Phi_\mu = \frac{N_{\mathrm{tracks}}}{A \cdot \mathcal{L}_{\mathrm{int}} \cdot \varepsilon}
    \label{eq:MuonFlux}
\end{equation}
where $N_{\mathrm{tracks}}$ is the number of muon tracks that traverse the effective detector area $A$ used for the measurement, $\mathcal{L}_{\mathrm{int}}$ is the integrated luminosity at IP1 and $\varepsilon$ is the tracking efficiency of the algorithm for the corresponding detector subsystem.

\section{Detector and coordinate system}
\label{sec:Detector}

\begin{figure*}[!bht]
    \centering
    \includegraphics[width=0.825\textwidth]{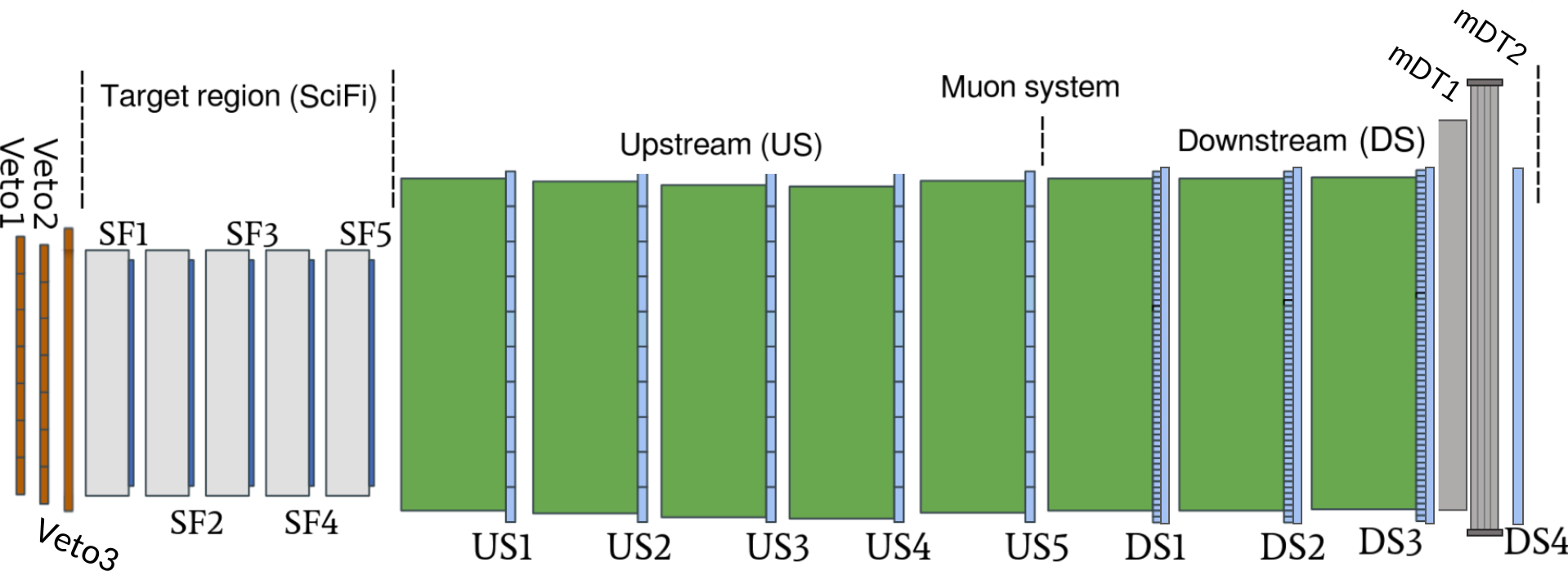}
    \caption{Schematic side-view (YZ plane) of the SND@LHC detector layout (not to scale). The setup includes the third veto plane (installed in 2024) and the two mini-drift tube (mDT) planes (added in 2025).}
    \label{fig:Detector}
\end{figure*}

The SND@LHC detector \cite{bib:detector} is a hybrid system featuring an $\sim$830 kg target formed of tungsten plates interleaved with
nuclear emulsion films and electronic trackers, which collectively function as an electromagnetic calorimeter (ECAL). The target is preceded by a veto system to tag charged particles originating from the IP1 direction, while a hadronic calorimeter (HCAL) and a muon system are situated downstream of the target (see \mbox{Figure \ref{fig:Detector}}).

Located in the TI18 tunnel, the detector is positioned approximately \mbox{480 m} from the ATLAS interaction point along the collision axis. This location is shielded by the bending of charged particles originating from IP1 by the LHC magnets and the attenuation of neutral hadrons by \mbox{100 m} of rock.

The veto system is made up of parallel planes of scintillator bars. Each layer consists of seven EJ-200 scintillator bars with dimensions of \mbox{$42 \times 6 \times 1$ cm$^3$} \\wrapped in aluminized mylar foil to prevent light leakage between adjacent bars. The horizontal scintillator bars are equipped with silicon photomultipliers (SiPMs) at both ends, whereas the vertical bars have SiPMs only at the top. At the start of 2024, the entire system was lowered via ground excavation to have better coverage of the target in a higher-intensity region of the neutrino flux (closer to the collision axis), and a third veto plane was installed to further improve its efficiency \cite{bib:3rd_veto_plane}. Since the veto system is not utilized for track reconstruction, these upgrades have no impact on the muon flux measurement.

The target, serving as the vertex detector, comprises five walls of Emulsion Cloud Chambers (ECCs). In this work, however, the analysis is restricted to data recorded by the electronic detectors.

Interleaved with the target walls are the Scintillating Fiber (SciFi) tracker planes, which provide precise measurements of particle positions and electromagnetic shower energies. Each plane consists of two orthogonal layers (vertical and horizontal). Every layer is composed of three fiber mats, each measuring \mbox{$13 \times 39$ $\mathrm{cm^2}$}. The scintillating fibers have a diameter of \mbox{$250 \; \mathrm{\upmu m}$} and are arranged in six staggered layers with a total thickness of \mbox{$1.35$ mm}. The readout system utilizes SiPM multichannel arrays. Each mat is read out by four SiPMs with 128 channels of width \mbox{$250 \; \upmu \mathrm{m}$}, totaling 512 channels per mat and 1536 channels per plane. For this analysis, the planes are labeled SF1 to SF5, ordered from most upstream to most downstream.

The muon system consists of eight iron blocks and nine scintillator planes arranged alternately along the beam direction. The iron blocks, each \mbox{$20$ cm} thick, serve as the absorber material for the HCAL. The scintillator planes are categorized into two types: upstream (US) and downstream (DS).

The US system comprises the first five planes following the target region, labeled US1 to US5 in order of increasing distance from the target. Each plane is composed of ten horizontal scintillator bars with dimensions of \mbox{$1 \times 6 \times 82.5$ cm$^3$}. These bars are read out by eight SiPMs at both ends, providing precise timestamps for traversing particles and measuring hadronic shower energy.

The DS system follows the US planes and offers finer granularity for muon identification and position measurement. DS planes consist of two orthogonal layers of scintillator bars (one horizontal and one vertical), except for the fourth plane (DS4), which contains only vertical bars. Each bar has a \mbox{$1 \times 1$ cm$^2$} cross-section, with lengths of \mbox{$82.5$ cm} for horizontal bars and \mbox{$63.5$ cm} for vertical bars. Horizontal bars are read out by a single SiPM at each end, while vertical bars are read out by a single SiPM at the top. Similarly to the SciFi, the DS planes are labeled DS1 to DS4.

Since 2025, the fourth DS plane has been relocated further downstream, with two drift tube planes (mDTs) installed between DS3 and DS4 to enhance muon tracking \cite{bib:miniDT}. As the mDT data was not incorporated into this analysis and the relocation of DS4 did not alter the DS tracking performance, these modifications have no effect on the muon flux measurement.

The system utilizes a triggerless data acquisition architecture, where all signals exceeding preset thresholds are processed by the front-end electronics and clustered in time to form events. An online software noise filter is subsequently applied. This maintains both a negligible detector deadtime and a negligible loss in signal efficiency. Events are recorded if they have coincident hits across multiple detector planes.

We employ a right-handed orthogonal coordinate system with its origin at \mbox{480 m} from IP1 along the LHC beam collision axis (assuming a null crossing angle). The $z$-axis is oriented along the detector's longitudinal axis, while the $y$-axis points upward.

\section{Data and Monte Carlo simulations}
\label{sec:DataAndSimulations}

\subsection{Data sample}
\label{sec:DataSample}

\begin{sloppypar}
Each dataset analyzed corresponds to a specific LHC fill. For this study, full track reconstruction was performed on data from five fills in 2023 and two in 2024, including both proton and heavy-ion collisions (specifically $^{208}\text{Pb}^{82+}$). While 2025 data were limited to representative subsamples, one proton and one heavy-ion fill are included to facilitate a direct comparison with the \mbox{2023–2024} datasets. This comparison is particularly relevant following the 2025 introduction of quasi-continuous crossing-angle anti-levelling, where the crossing angle is reduced as beam intensity decreases \cite{bib:lhc_highlights}. To provide a direct baseline for the heavy-ion measurements, a 2024 proton reference fill recorded at the reduced center-of-mass energy of \mbox{$\sqrt s = 5.36 \text{ TeV}$} is also included. All ten fills, including those from 2025 and the 2024 reference fill, are summarized in Table \ref{table:pp-PbPb-comparison}. The integrated luminosity for each LHC fill was provided by the ATLAS \mbox{Collaboration \cite{bib:AtlasLumi}}.
\end{sloppypar}

\begin{table*}[!hbt]
    \centering
    \caption{Beam attributes of heavy-ion and proton LHC fills used for muon flux measurement with high-statistics track reconstruction. The given values are the date of the fill number, the number of bunches per beam N, the crossing angle at IP1, the duration of stable beams operation of the fill and the integrated luminosities $\mathcal{L}_{\mathrm{int}}$. The crossing angle is given as a range for 2025 proton data due to the introduction of crossing-angle anti-levelling \cite{bib:lhc_highlights}, while fill 8695 lists the two specific values used. The uncertainty on the integrated luminosity is \mbox{3.5\%} for heavy ions and \mbox{2\%} for protons \cite{bib:AtlasLumi}. The center-of-mass energy is $\sqrt s = 13.6 \text{ TeV}$ and the time between bunches is \mbox{25 ns} for proton fills and \mbox{50 ns} for heavy-ion fills (except for the reference proton fill 10310).}
    \label{table:pp-PbPb-comparison}
    
    \begin{threeparttable}
    
    \begin{tabular}{c | c c c c c c} \hhline{=======}
          &   \multicolumn{5}{c}{\textbf{Heavy-Ion Collisions}} \\
          LHC Fill Number & Date & N & \makecell{IP1 Crossing Angle [$\upmu$rad]} & \makecell{Duration\ [h]} & $\mathcal{L}_{\mathrm{int}}$ [$\upmu$b$^{-1}$] \\ [0.5ex] \hline
        \makecell{9232} & 5 Oct 2023 & 1123 & $170$ & 11.0 & 61.01 \\ 
        \makecell{9291}  & 22 Oct 2023 & 1081 & $170$ & 6.5 & 56.98 \\ 
        \makecell{9312}  & 27 Oct 2023 & 960 & $170$ & 9.6 & 56.61 \\
        \makecell{9317}  & 29 Oct 2023 & 1080 & $170$ & 8.2 & 57.85 \\ 
        \makecell{10345} & 9 Nov 2024 & 1240 & $150$ & 6.5 & 65.49 \\
        \makecell{11318} & 20 Nov 2025 & 1240  & $150$  & 8.6  & 115.82  \\ \hhline{=======}
        
                    &   \multicolumn{5}{c}{\textbf{Proton Collisions}} \\
          LHC Fill Number & Date & N & \makecell{IP1 Crossing Angle [$\upmu$rad]} & \makecell{Duration\ [h]} & $\mathcal{L}_{\mathrm{int}}$ [$\upmu$b$^{-1}$] \\ [0.5ex] \hline
        \makecell{8695}  & 1 May 2023 & 911 & $-160$/$-135$ & 8.0 & 17.90 $\times$ 10$^{7}$ \\ 
        \makecell{9606}  & 7 May 2024 & 2352 & $150$ & 14.4 & 53.24 $\times$ 10$^{7}$ \\
        
        \makecell{10310\tnote{a}} & 2 Nov 2024 & 2352 & 170 & 22.9 & 8.82 $\times$ $10^7$  \\
        \makecell{10676}  & 31 May 2025 & 2460 & $160$-$120$ & 15.1 & 104.62 $\times$ $10^7$  \\
        \hhline{=======}
    \end{tabular}
    
    \begin{tablenotes}\footnotesize
        \item[a] Reference proton fill recorded at \mbox{$\sqrt{s} = 5.36 \text{ TeV}$}.
    \end{tablenotes}
    
    \end{threeparttable}
\end{table*}

\begin{sloppypar}
    To investigate the time dependence of the muon flux, a sampling approach was adopted, analyzing a subset of events from selected LHC fills throughout the data-taking period, including the entirety of the 2025 dataset. For these fills, track reconstruction was performed on a small fraction of the total events, with the results subsequently scaled to account for this subsampling. This approach is sufficient as the total uncertainty is dominated by systematic effects rather than statistics, as detailed in \mbox{Section \ref{sec:MuonFlux}}.

\end{sloppypar}

\subsection{Monte Carlo simulations}
\label{sec:Simulations}

\begin{sloppypar}
The simulation of the hadron collisions at the LHC interaction points is performed by CERN's SY-STI team with the FLUKA Monte Carlo simulation package and the embedded external event generator DPMJET \cite{bib:FLUKA,bib:DPMJET}. The resulting muons and charged hadrons are transported through the tunnel and rock to a virtual plane of dimensions \mbox{4.5 $\times$ 4 m$^2$} located $\sim$\mbox{35 m} upstream of the SND@LHC detector. Next, we use the output of this simulation to transport the particles to and through the detector with Geant4~\cite{bib:Geant4}.

The heavy-ion background was modeled by combining two distinct datasets: nuclear interactions (NI) from inelastic hadronic collisions and electromagnetic dissociation (EMD). These components are weighted according to their respective cross-sections to yield a total flux. So far, heavy-ion simulations were provided only for the 2022 LHC configuration.

\end{sloppypar}


\section{Muon tracking with the electronic detectors}
\label{sec:Tracks}

\subsection{Track finding}
\label{sec:TrackFinding}

The Hough transform algorithm~\cite{bib:HoughTransform} maps hits from the detector coordinate space into a binned parameter space (Hough space). In this space, each hit is represented by a curve, and the intersection of multiple curves within a single bin identifies a common trajectory. Conceptually, each hit casts a ``vote'' for every bin its curve passes through. The bins that accumulate the highest number of votes are then mapped back to physical space, effectively grouping hits that form a line to identify candidate tracks.

\subsection{Track fitting}
\label{sec:TrackBuilding}

\begin{sloppypar}
Once a track is identified, it is reconstructed using the Kalman filter as implemented in the Genfit package~\mbox{\cite{bib:Kalman,bib:Genfit}}. A tolerance parameter defines a window around the candidate line, potentially incorporating additional hits not identified in the initial Hough transform. At SND@LHC, track reconstruction is done separately in the XZ and YZ planes, with the combination of which a 3D track can be constructed. The primary downside of this 3D track construction mechanism is the ambiguity that arises in the pairing process when multiple tracks are present. For this analysis, only single track events are considered, while multi-muon events, arising either from particle decays at IP1 or from muon interactions in the upstream rock, are excluded. This is justified by their scarcity (approximately $1/10^{6}$ for two muons and even less for three) and the fact that they are the subject of separate studies.
\end{sloppypar}

Track reconstruction requires hits in at least three planes, both horizontal and vertical. The Hough transform typically produces multiple candidate tracks, with the final track chosen as the one containing the highest number of votes by the Hough transform algorithm. Since there are significant differences in the design and performance of the SciFi and DS components, tracking is done separately in the two subsystems.

\subsection{Tracking efficiency}
\label{sec:TrackingEfficiency}

\begin{sloppypar}
    Tracking efficiency is estimated by identifying a ``tagging track'' in either the SciFi or DS system that extrapolates into the acceptance of the other. Tracks found in the other system of tagged events are referred to as candidate tracks. A tag and candidate track pair are considered successfully matched if their extrapolated crossing points on a predefined reference plane between the two detectors lie within 3 cm of each other. The efficiency is then calculated as the ratio of matched pairs to the total number of tagging tracks.

\begin{figure}[!htbp]
    \centering
    \includegraphics[width=0.41125\textwidth]{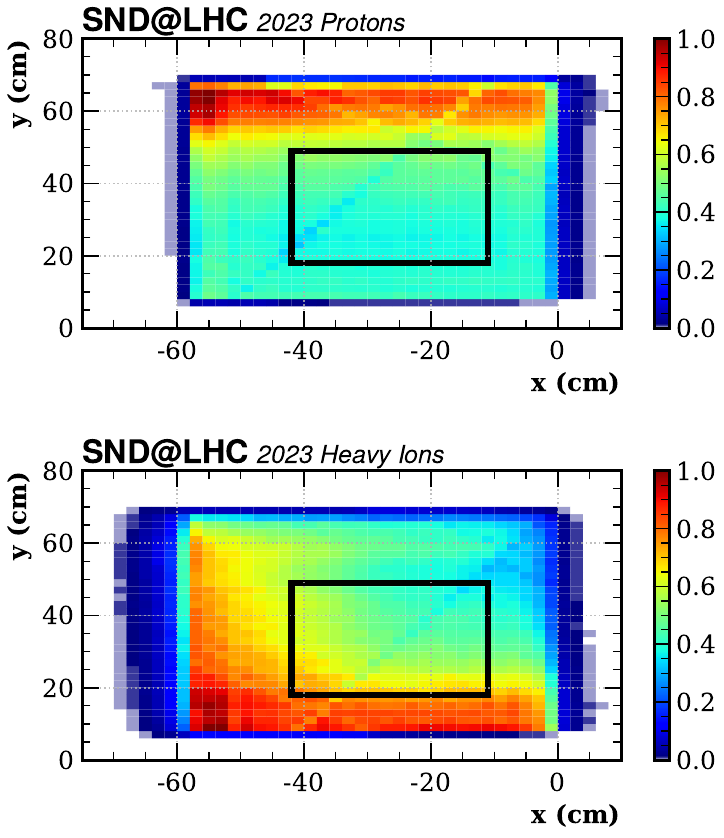}
    
    \caption{Normalized DS Hough transform track density distribution at \mbox{$z = 430$~cm} in data. Higher density is observed at the upper part for proton collisions and in the lower part for heavy-ion collisions. The box illustrates the fiducial region of uniform tracking efficiency for both SciFi and DS tracking. The diagonal lines of lower occupancy are artifacts of the 2D binning.}
    \label{fig:xyDistribution}
\end{figure}

\begin{figure}[!htbp]
    \centering
    \includegraphics[width=0.41125\textwidth]{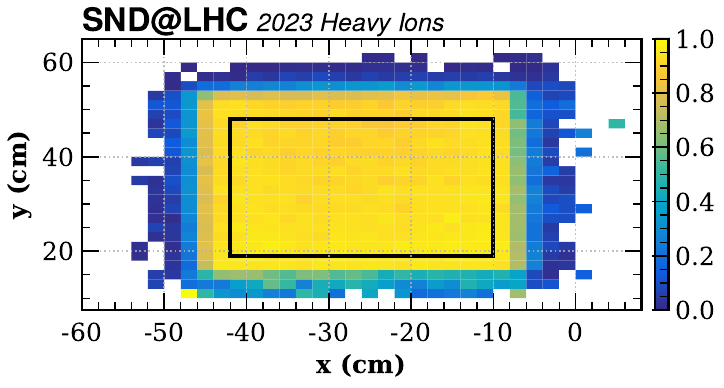}
    
    \caption{SciFi tracking efficiency distribution in data. The box illustrates the selected fiducial region.}
    \label{fig:trkeff-xy}
\end{figure}

    A fiducial region of uniform tracking efficiency is defined for both the SciFi and DS measurements. The DS track density distribution and the corresponding fiducial area are shown in \mbox{Figure \ref{fig:xyDistribution}}, while the resulting spatial map of tracking efficiency is illustrated in \mbox{Figure \ref{fig:trkeff-xy}}. The total efficiency is defined as the average across the fiducial area, where the contribution of each bin is weighted by its local track density. The calculated uncertainties include a global binomial statistical component and the spatial variance (systematic spread) across individual bins to address detector non-uniformity. These results are summarized in \mbox{Table \ref{tab:TrackingEfficiencies}}.
\end{sloppypar}

\begin{table}[!ht]
    \caption{Comparison of tracking efficiency across the two detector subsystems. Fluctuations between years and bunch types (protons vs. heavy ions) reflect variations in the muon energy spectra and the energy-dependent nature of tracking performance. Quoted uncertainties include both statistical and systematic components. The systematic uncertainty is derived from the efficiency variance across the fiducial region, while the statistical component is negligible, contributing less than 0.1\% to the total uncertainty.}
    \label{tab:TrackingEfficiencies}
    \centering
    \setlength{\tabcolsep}{3pt}
    \begin{tabular}{l | c c} 
        \toprule
         & \textbf{SciFi Efficiency} & \textbf{DS Efficiency} \\
        \hhline{===}
        \textit{2023 Data} \\
        \quad Protons      & 0.949 ± 0.009 & 0.932 ± 0.012 \\
        \quad Heavy Ions   & 0.929 ± 0.018 & 0.894 ± 0.029 \\
        \midrule
        \textit{2024 Data} \\
        \quad Protons      & 0.911 ± 0.011 & 0.922 ± 0.013 \\
        \quad Heavy Ions   & 0.857 ± 0.016 & 0.853 ± 0.021 \\
        \midrule
        \textit{2025 Data} \\
        \quad Protons     & 0.922 ± 0.008 & 0.924 ± 0.014 \\
        \quad Heavy Ions  & 0.898 ± 0.018 & 0.881 ± 0.031 \\
        \bottomrule
    \end{tabular}
\end{table}

\section{Angular distribution}
\label{sec:AngularDistribution}

\begin{figure*}[ht!]

    \centering
    \includegraphics[width=1\textwidth]{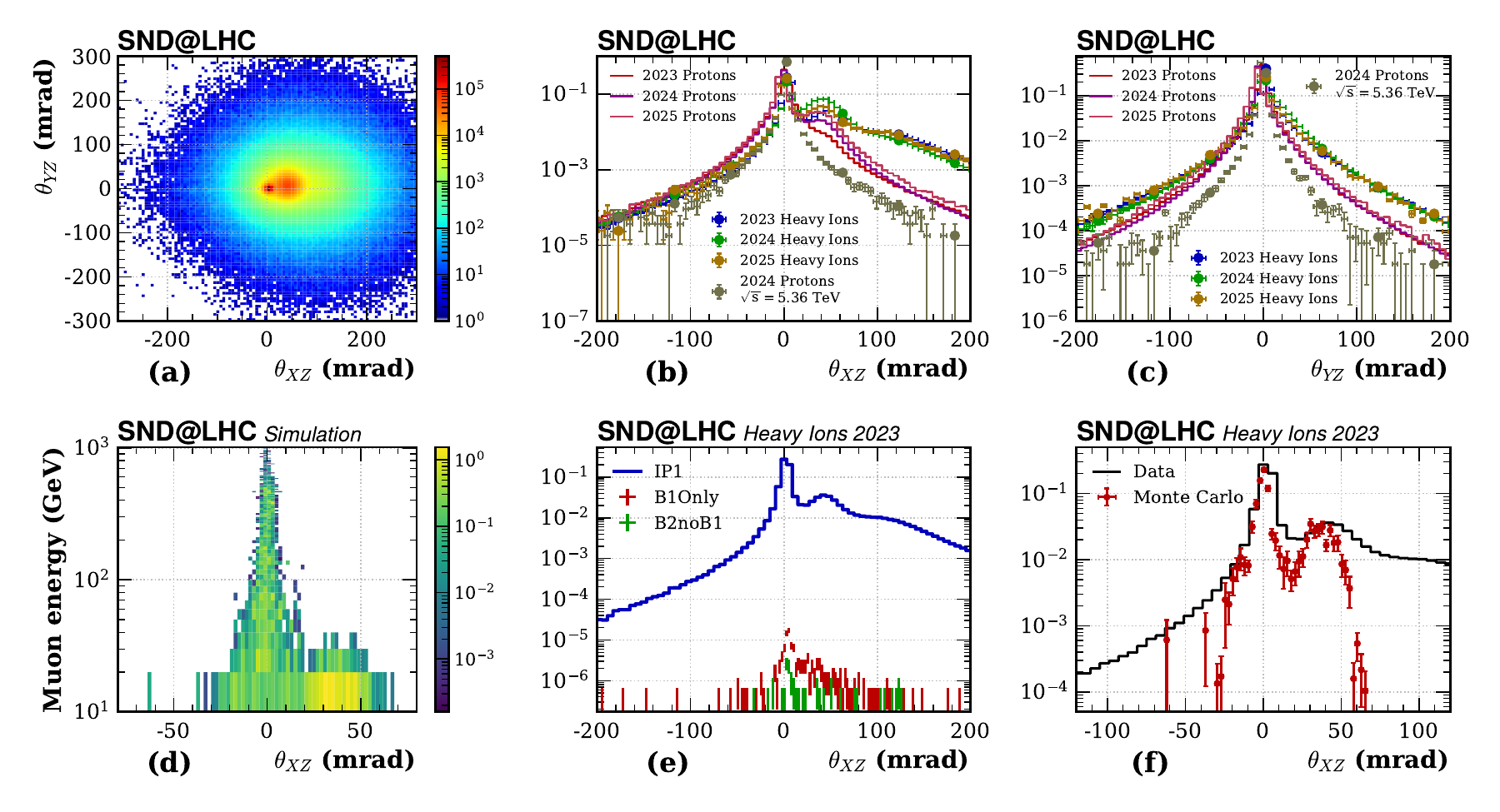}
    \caption{Track angle distributions and source separation. Panel \textbf{(a)} is in total counts, while panels \textbf{(b, c, e, f)} are normalized to unit area. Panel \textbf{(d)} shows a 2D histogram of track angle versus muon energy for heavy-ion simulations in arbitrary units. The labels IP1, B1Only, and B2noB1 denote events coincident with an IP1 collision, a non-colliding beam 1, or a non-colliding beam 2 bunch slot, respectively. Contributions from non-colliding beams are negligible and are excluded from the final muon flux results.}
    \label{fig:Angles}
\end{figure*}

Given the detector's considerable distance from IP1, it is anticipated that most tracks will exhibit minimal slopes. However, the angular distribution in \mbox{Figure \ref{fig:Angles}} shows a secondary peak at \mbox{$\theta_{XZ} \approx 45$ mrad} in the horizontal projection, along with a plateau near \mbox{110 mrad}. Monte Carlo simulations indicate that this secondary peak originates from low-energy muons produced via hadron decays, which are more abundant in heavy-ion data.

\begin{figure}[ht!]

    \centering
    \includegraphics[width=0.39\textwidth]{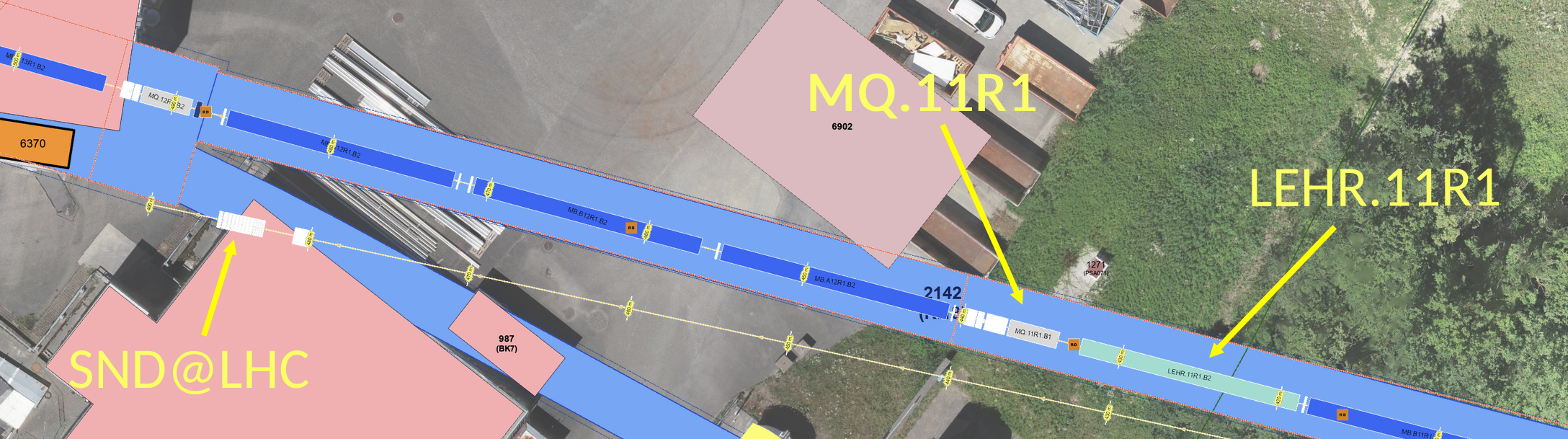}
    \caption{The LHC elements LEHR.11R1 and MQ.11R1. Pions and kaons, generated through ion interactions at LEHR.11R1, decay into muons. These muons are subsequently directed toward the detector by the quadrupole magnet MQ.11R1 and following dipoles.}
    \label{fig:MuonDeflection}
\end{figure}

\begin{figure}[ht!]
    \centering
    \includegraphics[width=0.39\textwidth]{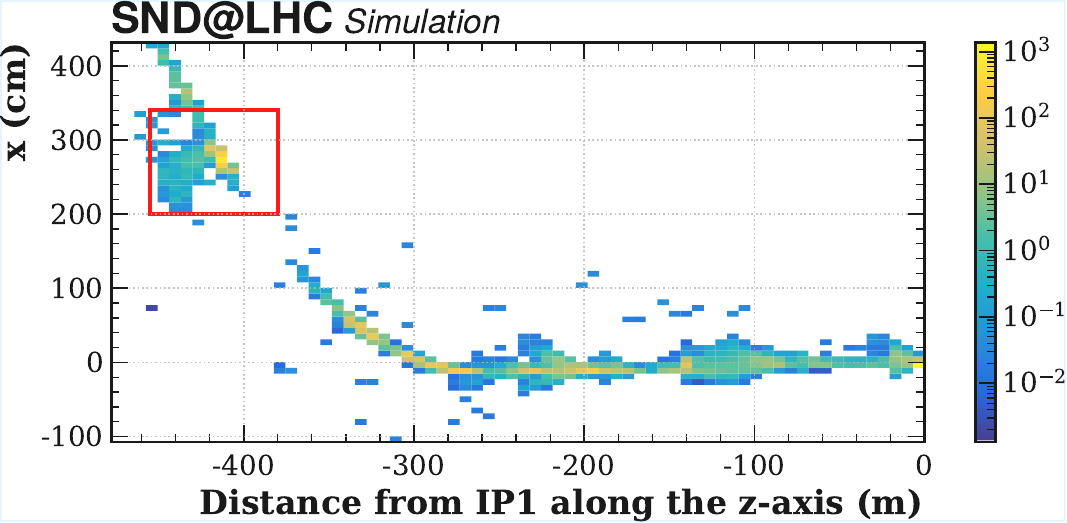}
    \caption{Muon production distribution along the beamline in heavy-ion simulations. The coordinate system is centered at IP1 with the \mbox{$z$-axis} pointing along the beam collision axis, away from SND@LHC. Concentration of muon production within the boxed area, approximately \mbox{415 m} from IP1, aligns with the position of the LHC \mbox{half-cell 11}.}
    \label{fig:lastdec}
\end{figure}

\begin{sloppypar}
A positive angle corresponds to muons traveling away from the LHC's center. Extrapolating from the detector to its intersection with the LHC ring places the deflection or production source approximately \mbox{60 m} upstream. The GIS portal map \cite{bib:CERN_GIS} identifies two LHC elements at this distance: the quadrupole magnet MQ.11R1 and the empty cryostat LEHR.11R1, as shown on \mbox{Figure \ref{fig:MuonDeflection}}. FLUKA simulations provide insights into the muon origin before reaching the virtual plane. Initially, pions and kaons are generated by heavy-ion interactions at the LEHR.11R1 site. Subsequently, these particles decay into muons, whose trajectories are directed to the detector by the quadrupole magnet MQ.11R1 and following dipoles. The muon production points in the simulation are given in \mbox{Figure \ref{fig:lastdec}}.
\end{sloppypar}

\subsection{Muon energy spectrum discrepancies}
\label{sec:muon_energies}

Monte Carlo simulations indicate that muons originating from heavy-ion collisions exhibit a notably softer energy spectrum compared to those from proton collisions. This spectral softening may be attributed to two primary factors, the lower center-of-mass energy per nucleon pair in heavy-ion fills (\mbox{$\sqrt{s_\text{NN}} = 5.36$ TeV} vs. \mbox{$13.6$ TeV} for protons) and the contribution of EMD processes characteristic to heavy-ion interactions.

\begin{figure}[h!]
    \centering
    \includegraphics[width=0.42\textwidth]{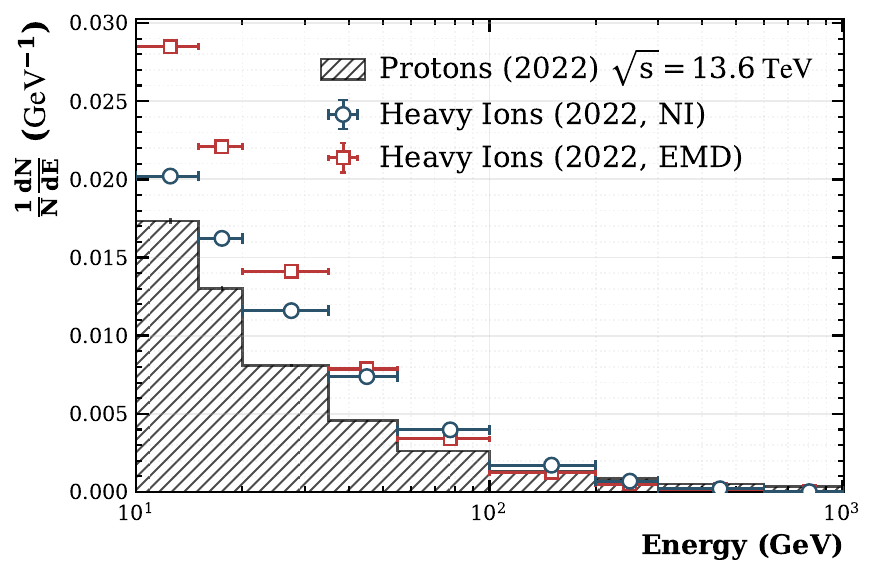}
    \caption{Simulated muon energy spectra at the SND@LHC detector location. The plot shows the independent spectral contributions from EMD and NI, alongside the 2022 proton collision baseline. Histograms are normalized to unit area and corrected for bin width to show the probability density.}
    \label{fig:emd_vs_ni}
\end{figure}

As shown in \mbox{Figure \ref{fig:emd_vs_ni}}, a decomposition of the heavy-ion simulation reveals that while both NI and EMD components are softer than the \mbox{$13.6$ TeV} proton baseline, the EMD process specifically populates the lowest energy bins.

\begin{sloppypar}
   While proton simulations at a matching \mbox{$\sqrt{s} = 5.36$ TeV} are currently unavailable, the 2024 reference dataset (Fill 10310) provides a baseline closer to the heavy-ion collision energy. The narrower angular distribution observed in this reference fill indicates a higher mean muon momentum compared to the heavy-ion data. This suggests that the energy difference is not solely a function of the collision energy, but is likely driven by the EMD contribution inherent to heavy-ion collisions. However, a definitive conclusion cannot be made without a validation with energy-matched simulations to isolate these physical processes from other confounding experimental variables. 
\end{sloppypar}

\FloatBarrier

\section{Muon flux}
\label{sec:MuonFlux}

\begin{sloppypar}
    The muon flux per reaction is different across the years due to changing LHC configurations. The difference between proton and heavy-ion collisions is an interplay of many factors including different interaction mechanisms and also additional tuning of the machine for ion operation.
\end{sloppypar}

\begin{figure*}[!t]
    \includegraphics[width=0.9\textwidth]{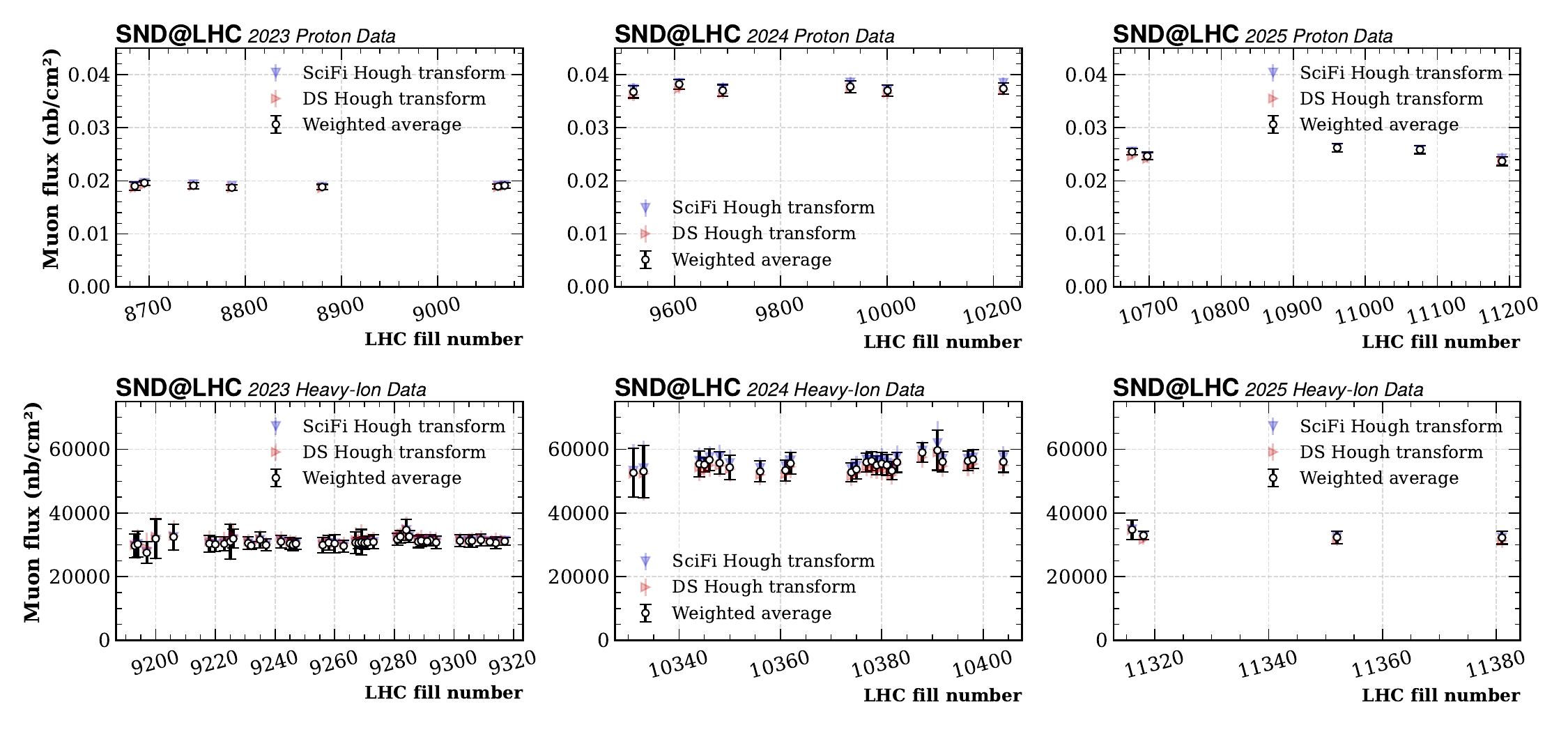}
    \centering
    
    \caption{Muon flux measurement for all LHC fills with production-mode filling schemes. Vertical error bars represent the total uncertainty, comprising both statistical and systematic components.}

    \label{fig:FluxVsRuns}
\end{figure*}

\begin{figure}[!ht]
    \includegraphics[width=0.49\textwidth]{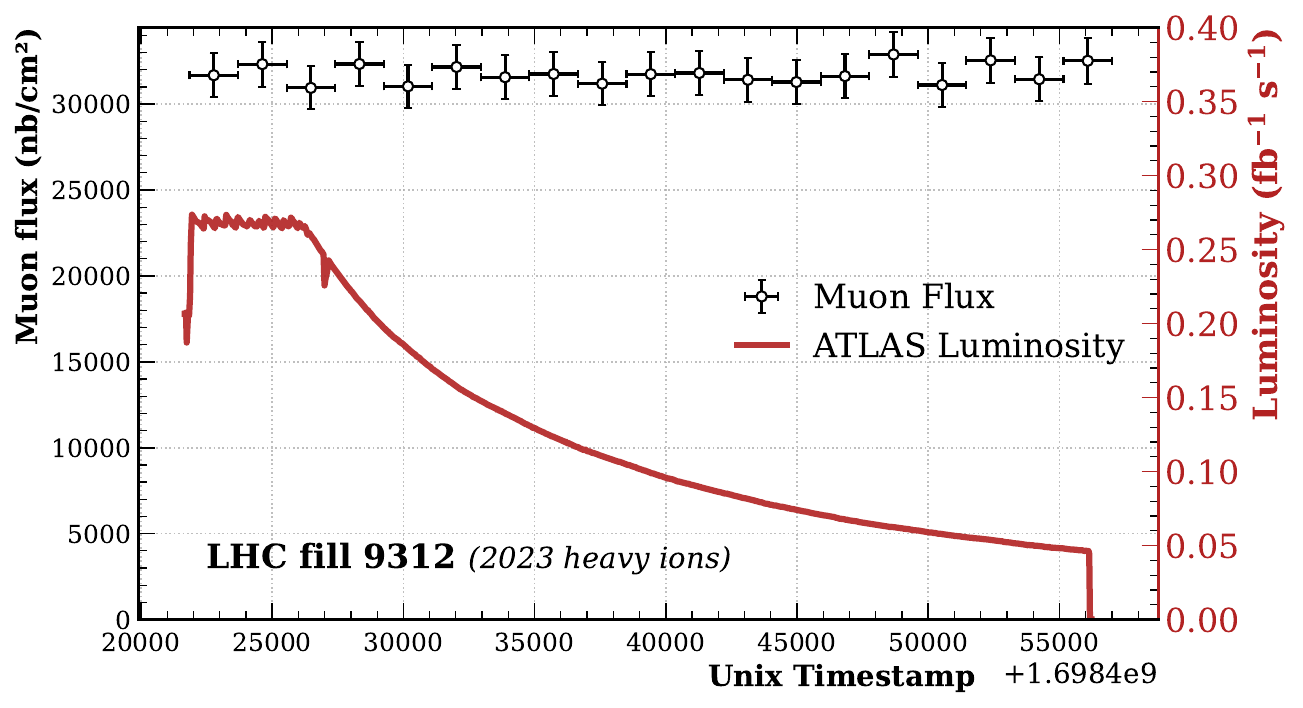}
    \centering
    
    \caption{Evolution of the muon flux within a representative LHC fill, calculated over consecutive time intervals.}

    \label{fig:flux_vs_lumi_evolution}
\end{figure}

\begin{sloppypar}
    The fiducial area for both subsystems has been chosen with boundaries of \mbox{$-42$ cm $\leq$ X $\leq$  $-11$ cm} and \mbox{$18$ cm $\leq$ Y $\leq$ $49$ cm}. Muon flux contributions from non-IP1 meson decays originating \mbox{60 m} upstream of the detector cannot be isolated and are thus included in the final results. All independent measurements from the two subsystems, along with the weighted average, are presented in \mbox{Figure \ref{fig:FluxVsRuns}}, which shows that the flux remains stable over time. This temporal stability within individual LHC fills is further confirmed in \mbox{Figure \ref{fig:flux_vs_lumi_evolution}}, demonstrating that the measurement is independent of the instantaneous luminosity during both the initial leveling and subsequent non-leveling phases of beam operation.
\end{sloppypar}

\begin{sloppypar}
  The muon flux results for heavy-ion and proton data from 2023 to 2025 are summarized in \mbox{Table \ref{tab:MuonFlux}}. No heavy-ion data was collected in 2022 due to the decision to conclude the run two weeks early to mitigate the European energy crisis \cite{bib:lhc_2022}. Heavy-ion collision simulations were only provided with the 2022 machine configuration.  
\end{sloppypar}

\begin{sloppypar}
    For proton data, the observed flux deviates from predictions by 10\% to 20\%. Given the complexity of the simulations, from the collision generator to the propagation through rock of varying composition, with cumulative uncertainties at each stage, this level of agreement can be considered excellent.
\end{sloppypar}

\begin{table*}[!ht] 
    \centering
    \caption{Measured and simulated muon fluxes for the 2023–2025 periods. One 2024 proton reference fill recorded at a reduced center-of-mass energy of $\sqrt{s} = 5.36 \text{ TeV}$ is also included. Relative contributions to the total uncertainty are provided for each measurement. For these calculations, a luminosity uncertainty of 2.0\% for proton data and 3.5\% for heavy-ion data was adopted \cite{bib:AtlasLumi}.}
    \label{tab:MuonFlux}
    \small 
    \begin{tabular*}{\textwidth}{@{\extracolsep{\fill}} l c c c c @{}} 
    \toprule
    & & \multicolumn{3}{c}{Uncertainty Breakdown (\% variance)} \\ \cmidrule{3-5}
    Dataset & Muon Flux & Statistics & Tracking Efficiency & Luminosity \\ \midrule 

    \multicolumn{5}{l}{\textit{Heavy-Ion Collisions ($10^4$ nb/cm$^2$)}} \\ \midrule
    2023 Data & $3.11 \pm 0.12$ & $<0.01$ & 22.06 & 77.94 \\
    2024 Data & $5.53 \pm 0.22$ & $<0.01$ & 22.80 & 77.20 \\
    2025 Data & $3.24 \pm 0.13$ & 0.06 & 20.48 & 79.46 \\ \addlinespace
    2022 Simulation & $2.99 \pm 0.09$ & 96.70 & 3.30 & -- \\ \midrule

    \multicolumn{5}{l}{\textit{Proton Collisions ($10^{-2}$ nb/cm$^2$)}} \\ \midrule
    
    2023 Data        & $1.90 \pm 0.04$ & 0.02 & 16.94 & 83.04 \\
    2023 Simulation  & $1.67 \pm 0.05$ & 94.71 & 5.29 & -- \\ \addlinespace
    
    2024 Data        & $3.76 \pm 0.09$ & 0.01 & 27.62 & 72.37 \\
    2024 Data (Reference Fill)\textsuperscript{a} & $4.21 \pm 0.14$ & 1.33 & 64.50 & 34.16 \\ 
    2024 Simulation  & $3.34 \pm 0.12$ & 87.88 & 12.12 & -- \\ \addlinespace
    
    2025 Data        & $2.48 \pm 0.05$ & 0.03 & 12.64 & 87.33 \\ 
    2025 Simulation  & $3.13 \pm 0.14$ & 87.21 & 12.79 & -- \\ \midrule

    2022 Data\textsuperscript{b} & $2.06 \pm 0.12$ & 8.33 & \multicolumn{2}{c}{91.67 (Combined Systematics)} \\ \bottomrule
    \end{tabular*}

    \vspace{1ex}
    \raggedright
    \footnotesize
    \textsuperscript{a} LHC fill 10310, measured at $\sqrt{s} = 5.36 \text{ TeV}$. \\
    \textsuperscript{b} 2022 proton flux results are taken from Ref.~\cite{bib:ppMuonFlux}.
\end{table*}

\subsection{LHC configurations}
\label{sec:LHC}

The significant increase in muon flux by a factor of $\sim$2 between 2023 and 2024 can be attributed to changes in the LHC \mbox{configuration \cite{bib:lhc_settings_2023_and_2024}}. Most notably, the LHC IR1 triplet polarity was reversed in 2024, inverting the quadrupole sequence from Focus-Defocus-Focus (FDF) to Defocus-Focus-Defocus (DFD) to optimize the machine lifetime and Run 3 performance. For heavy-ion operations, the amplitude of the bound-free pair production (BFPP) orbit bumps was also adjusted to prevent magnet quenching under these new settings. As a result, an increased number of TeV-energy muons reached the detector.

In 2025, the nominal settings were restored, but the crossing angle at IP1 was set to the horizontal plane. This configuration resulted in a muon flux $\sim$23\% higher than in 2023 for proton collisions, whereas no such increase was observed for heavy-ion data.
The detailed analysis is part of a separate study and is not addressed here.

A notable observation is the elevated muon flux and narrowed angular distribution measured for the low-energy reference proton fill compared to standard 2024 proton data. This discrepancy is likely attributable to two key machine configuration differences. First, the muon background is sensitive to the settings of the physics debris collimators (TCLs), located approximately 270 m from IP1 in the direction of the SND@LHC detector. During the reference fills, the TCL half-gaps were set to 25 mm, significantly wider than the standard 2024 operating parameters of \mbox{12.3 mm}, \mbox{14.5 mm}, and \mbox{1.586 mm} for TCL4, TCL5, and TCL6, respectively. These wider gaps permit a greater flux of forward pions and kaons to propagate and subsequently decay into the muons detected by the experiment. Second, the reference fill utilized de-squeezed optics with an amplitude function $\beta^*$ of \mbox{3.0 m}, in contrast to the \mbox{0.5 m} employed for standard \mbox{13.6 TeV} fills. The resulting less aggressive magnetic corrections likely lead to reduced proton losses at the dispersion suppressor near LHC \mbox{half-cell 11}. A full quantitative assessment of how the increased $\beta^*$ affects the dispersion function has not yet been established.


\section{Conclusion}
\label{sec:Conclusion}

We have reported the muon flux measurements for heavy-ion and proton collisions at the SND@LHC experiment using data collected between 2023 and 2025. Muons originating from beam interactions at LEHR.11R1 follow the same trajectory as IP1 muons and are synchronous with collision events. Consequently, they cannot be temporally separated and are included in the reported flux values. 

The measured muon fluxes for heavy-ion collisions are:
\begin{subequations}
    \begin{align}
        \textbf{2023 Data:} &\quad \Phi_{\mu} = (3.11 \pm 0.12) \times 10^4 \;\mathrm{nb/cm}^2 \\
        \textbf{2024 Data:} &\quad \Phi_{\mu} = (5.53 \pm 0.22) \times 10^4 \;\mathrm{nb/cm}^2 \\
        \textbf{2025 Data:} &\quad \Phi_{\mu} = (3.24 \pm 0.13) \times 10^4 \;\mathrm{nb/cm}^2
    \end{align}
    \label{eq:MuonFluxResultPbPb}
\end{subequations}
The measured muon fluxes for proton data are:
\begin{subequations}
    \begin{align}
        \textbf{2023 Data:} &\quad \Phi_{\mu} = (1.90 \pm 0.04) \times 10^{-2} \; \mathrm{nb/cm}^2 \\
        \textbf{2024 Data:} &\quad \Phi_{\mu} = (3.76 \pm 0.09) \times 10^{-2} \; \mathrm{nb/cm}^2 \\
        \textbf{2025 Data:} &\quad \Phi_{\mu} = (2.48 \pm 0.05) \times 10^{-2} \; \mathrm{nb/cm}^2
    \end{align}
    \label{eq:MuonFluxResultPP}
\end{subequations}

Systematic uncertainties dominate, with the statistical component contributing less than 0.1\% for both proton and heavy-ion data (see \mbox{Table \ref{tab:MuonFlux}}). Unlike the 2022 proton data analysis, a unified fiducial area was adopted for both SciFi and DS measurements. This consistency eliminates the choice of detector subsystem as a source of systematic uncertainty.

A comparison with Monte Carlo simulations shows agreement within $\sim$10\% for the 2023 and 2024 proton data, and within $\sim$20\% for the 2025 data. Heavy-ion simulations were performed using the 2022 LHC configuration. Given the similarities between the 2022 and 2023 heavy-ion LHC configurations, these simulations show good agreement ($\sim$5\%) with the 2023 experimental results.

A dedicated proton reference dataset at reduced collision energy was recorded in 2024 to provide a controlled baseline for heavy-ion measurements. Comparing this reference dataset to heavy-ion data at an equivalent energy, the narrower angular distribution observed in the reference fill indicates a higher mean muon momentum. When contrasted with the softer spectrum predicted by EMD processes characteristic of heavy-ion collisions, these observations suggest that EMD is the primary driver of the lower energy spectrum in the ion data. However, definitive confirmation of this hypothesis requires validation through energy-matched simulations.

Distinct from the ion-to-proton comparison, the reference fill also highlights the detector's sensitivity to machine-induced backgrounds. Compared to standard 2024 proton runs, the reference fill exhibited an \mbox{$\sim$11\%} increase in muon flux and a narrower angular distribution lacking the characteristic secondary peak at $\sim$45 mrad, which has been traced to proton losses at the dispersion suppressor near \mbox{half-cell 11}. These discrepancies are directly attributable to the specific machine configuration of the reference fill. The wider half-gaps of the TCLs facilitated an increased yield of forward-propagating pions and kaons, while the use of de-squeezed optics likely mitigated the beam losses at the dispersion suppressor.


\bibliographystyle{spphys}       
\bibliography{references}
\input{acknowledgments}
\clearpage
\input{authorlist}
\end{document}

%% file: acknowledgments.tex
\begin{acknowledgements}

\begin{sloppypar}
We acknowledge the support for the construction and operation of the SND@LHC detector provided by the following funding agencies:  CERN;  the Bulgarian Ministry of Education and Science within the National
Roadmap for Research Infrastructures 2020–2027 (object CERN) and for the support of this study under the National Program ``Young Researchers and Postdoctoral Students - 2''; ANID FONDECYT grants No. 3230806, No. 1240066, 1240216 and ANID  - Millenium Science Initiative Program -  $\rm{ICN}2019\_044$ (Chile); the Deutsche Forschungsgemeinschaft (DFG, ID 496466340); the Italian National Institute for Nuclear Physics (INFN); JSPS, MEXT, the~Global COE program of Nagoya University, the~Promotion and Mutual Aid Corporation for Private Schools of Japan for Japan; the National Research Foundation of Korea with grant numbers 2021R1A2C2011003, 2020R1A2C1099546, 2021R1F1A1061717, and 2022R1A2C100505; Fundação para a Ciência e a Tecnologia, FCT grant numbers  CEECIND/01334/2018, CEECINST/00032/2021 and PRT/BD/153351/2021 (Portugal), CERN/FIS-INS/0028/2021; the Swiss National Science Foundation (SNSF); TENMAK for Turkey (Grant No. 2022TENMAK(CERN) A5.H3.F2-1). J.C.~Helo~Herrera and O.~J.~Soto~Sandoval acknowledge support from ANID FONDECYT grants No.1241685 and 1241803. M.~Climescu, H.~Lacker and R.~Wanke are funded by the Deutsche Forschungsgemeinschaft (DFG, German Research Foundation), Project 496466340. This research was financially supported by the Italian Ministry of University and Research within the Prin 2022 program.
\end{sloppypar}

We express our gratitude to our colleagues in the CERN accelerator departments for the excellent performance of the LHC. We thank the technical and administrative staff at CERN and at other SND@LHC institutes for their contributions to the success of the SND@LHC efforts. We are particularly grateful to the CERN SY-STI team for their expertise regarding beam losses and for providing dedicated simulations of the orbit bump tests, which were critical to understanding the background conditions. We also thank our colleagues in the CERN BE-ABP group for the insightful discussions on LHC configurations. We thank Luis Lopes, Jakob Paul Schmidt and Maik Daniels for their help during the~construction.

\end{acknowledgements}

%% file: authorlist.tex
\onecolumn
\begin{center}
\textbf{The SND@LHC Collaboration}
\vspace{0.25cm}
\break
\author{D.~Abbaneo$^{9}$\orcidlink{0000-0001-9416-1742}},
\author{S.~Ahmad$^{42}$\orcidlink{0000-0001-8236-6134}},
\author{R.~Albanese$^{1,2}$\orcidlink{0000-0003-4586-8068}},
\author{A.~Alexandrov$^{1}$\orcidlink{0000-0002-1813-1485}},
\author{F.~Alicante$^{1,2}$\orcidlink{0009-0003-3240-830X}},
\author{F.~Aloschi$^{1,2}$\orcidlink{0000-0002-2501-7525}},
\author{K.~Androsov$^{6}$\orcidlink{0000-0003-2694-6542}},
\author{A.~Anokhina$^{3}$\orcidlink{0000-0002-4654-4535}},
\author{L.G.~Arellano$^{1,2}$\orcidlink{0000-0002-1093-1824}},
\author{C.~Asawatangtrakuldee$^{38}$\orcidlink{0000-0003-2234-7219}},
% ...
\end{center}

\begin{flushleft}
\begin{footnotesize}
$^{1}$Sezione INFN di Napoli, Napoli, 80126, Italy\linebreak
$^{2}$Universit\`{a} di Napoli ``Federico II'', Napoli, 80126, Italy\linebreak
$^{3}$Affiliated with an institute formerly covered by a cooperation agreement with CERN\linebreak
% ...
$^{6}$Institute of Physics, EPFL, Lausanne, 1015, Switzerland\linebreak
% ...
$^{9}$European Organization for Nuclear Research (CERN), Geneva, 1211, Switzerland\linebreak
% ...
$^{38}$Department of Physics, Faculty of Science, Chulalongkorn University, Bangkok, 10330, Thailand\linebreak
% ...
$^{42}$Affiliated with Pakistan Institute of Nuclear Science and Technology (PINSTECH), Nilore, 45650, Islamabad, Pakistan
% ...
\end{footnotesize}
\end{flushleft}